\documentclass[conference,compsoc]{IEEEtran}
\ifCLASSOPTIONcompsoc
  \usepackage[nocompress]{cite}
\else
  \usepackage{cite}
\fi
\usepackage{subfig}

\ifCLASSINFOpdf
  \usepackage[pdftex]{graphicx}
  \DeclareGraphicsExtensions{.pdf,.jpeg,.png}
\else
  \usepackage[dvips]{graphicx}
  \graphicspath{{../eps/}}
  \DeclareGraphicsExtensions{.eps}
\fi
\usepackage{amsmath}
\usepackage{algorithmic}

\usepackage{booktabs}

\usepackage{array}

\usepackage{stfloats}
\usepackage{url}

\usepackage{xcolor}

\usepackage{enumitem}

\usepackage[norule]{footmisc}

\providecommand{\keywords}[1]
{
{\small{\textbf{\textit{Keywords---}} #1}}
}

\IEEEoverridecommandlockouts
\begin{document}
%
\title{\vspace{-1em}A New MRAM-based Process In-Memory Accelerator for Efficient Neural Network Training with Floating Point Precision\vspace{-1em}}

\author{
\IEEEauthorblockN{Hongjie Wang\thanks{\textsuperscript{${\ast}$}\emph{Hongjie Wang and Yang Zhao contributed equally to this work.}}\textsuperscript{${\ast}$}, Yang Zhao\textsuperscript{${\ast}$}, Chaojian Li, Yue Wang, Yingyan Lin}
\IEEEauthorblockA{Department of Electrical and Computer Engineering, Rice University, TX, USA}
\vspace{-2em}
}


\maketitle

\indent\begin{abstract}

The excellent performance of modern deep neural networks (DNNs) comes at an often prohibitive training cost, limiting the rapid development of DNN innovations and raising various environmental concerns. To reduce the dominant data movement cost of training, process in-memory (PIM) has emerged as a promising solution as it alleviates the need to access DNN weights. However, state-of-the-art PIM DNN training accelerators employ either analog/mixed signal computing which has limited precision or digital computing based on a memory technology that supports limited logic functions and thus requires complicated procedure to realize floating point computation.   
In this paper, we propose a spin orbit torque magnetic random access memory (SOT-MRAM) based digital PIM accelerator that supports floating point precision. Specifically, 
this new accelerator features an innovative (1) SOT-MRAM cell, (2) full addition design, and (3) floating point computation. Experiment results show that the proposed SOT-MRAM PIM based DNN training accelerator can achieve 3.3$\times$, 1.8$\times$, and 2.5$\times$ improvement in terms of energy, latency, and area, respectively, compared with a state-of-the-art PIM based DNN training accelerator. 

\end{abstract}


%
\IEEEpeerreviewmaketitle

\vspace{-0.8em}
\section{Introduction}
\vspace{-1em}
The recent record-breaking predictive performance achieved by deep neural networks (DNNs) motivates a tremendously growing demand to bring DNN-powered intelligence into numerous applications. However, the excellent performance of modern DNNs comes at an often prohibitive training cost due to the vast volume of training data and model parameters required. For example, one forward pass of ResNet50 \cite{he2016deep} requires 4 GFLOPs (FLOPs: floating point operations per second) of computation and training requires $10^{18}$ FLOPs, which takes 14 days on one state-of-the-art NVIDIA M40 GPU \cite{you2018imagenet}.  As a result, training a state-of-the-art DNN model often demands considerable energy, along with the associated financial and environmental costs. 

To address the time and energy dominant data movements of DNNs, process in-memory (PIM) based accelerators have emerged as a promising solution. While various PIM based accelerators have been developed for efficient DNN inference \cite{ISAAC, PRIME, PipeLayer}, PIM based accelerators for efficient training is less explored. Furthermore, state-of-the-art PIM based DNN training accelerators use either analog/mixed signal computing, which has limited precision \cite{PipeLayer}, or digital computing using a memory technology that supports limited logic functions and thus requires complicated procedures to realize floating point computation \cite{FloatPIM}.


In this paper, we propose a spin orbit torque magnetic random access memory (SOT-MRAM)  based PIM accelerator that supports floating point computations which is often desirable in DNN training and features improved energy, time and area efficiency over state-of-the-art PIM accelerators for DNN training. 
The contribution of this paper can be summarized as follows: 
\vspace{-0.2em}
\begin{itemize}[leftmargin=*]

    \item We develop a 1T-1R SOT-MRAM memory cell which features an improved balance between computation flexibility and memory density, favoring more efficient PIM accelerators.
    \item We propose a new full addition (FA) design that requires fewer computing steps to finish addition operations compared with the state-of-the-art design in \cite{FloatPIM}.
    \item We develop efficient designs (reduced latency and energy) of the dominant floating point addition and multiplication for digital PIM accelerators, where the required latency and energy cost are analytically formulated.
    \item We integrate the aforementioned 1T-1R SOT-MRAM memory cell, new FA design, and efficient floating point computation to demonstrate a new SOT-MRAM PIM based DNN training accelerator. 
    
\end{itemize}
\vspace{-1.2em}
\section{Related Works and Background}\label{background}
\begin{figure}[!t]
\centering
\includegraphics[width=3.3in]{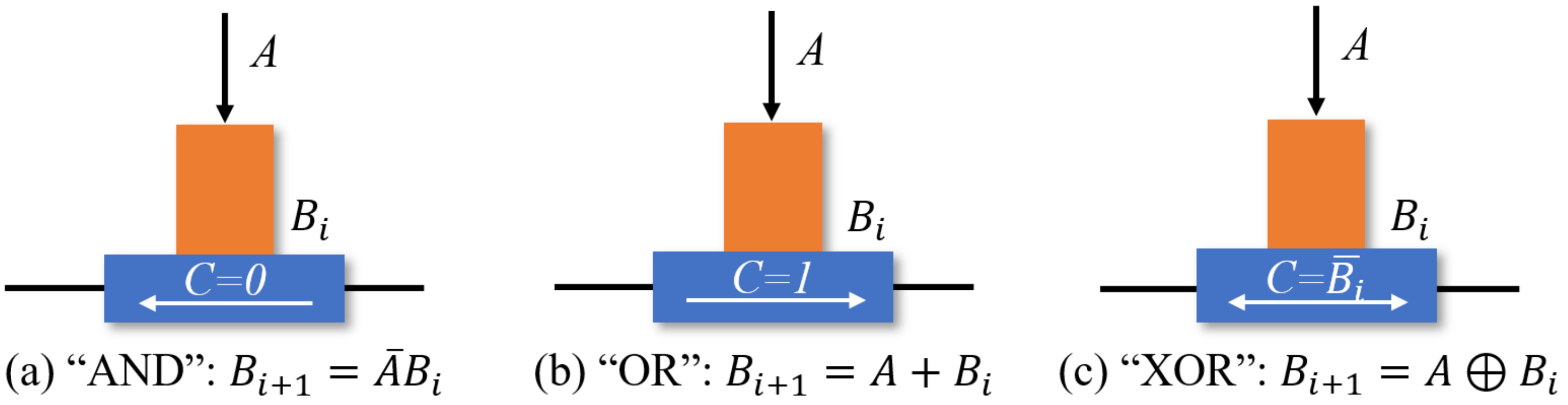}
\vspace{-0.3em}
\caption{
An implementation of (a) AND, (b) OR, and (c) XOR logic functions using a single MTJ device \cite{SOT-MRAMbasic}, where $A$ denotes the applied voltage $V_b$ (either logic 1 (e.g., $V_b=600$mV) or 0 (e.g., $V_b=0$V)); $B_i$ represents the initial resistance state with either a high (logic 1) or low resistance (logic 0); $C$ denotes the direction of the device's writing current; and $B_{i+1}$ represents the computing result.}
\label{fig_logicfunc}
\vspace{-1.5em}
\end{figure}

\vspace{-0.8em}
\textbf{SOT-MRAM based Logic Functions.} 
SOT-MRAM is a type of non-volatile memory that stores data via resistance state (high or low) of Magnetic Tunnel Junctions (MTJ). 
Compared with other memory technologies, SOT-MRAM has advantages of (1) high memory cell density, (2) potentially infinite endurance, and (3) requiring low memory writing current\cite{MRAMparameters}, making it an attractive candidate for PIM accelerators. There have been several works exploring SOT-MRAM based PIM accelerators for DNN inference. The works \cite{multilevelBCNN, PXNORBNN} propose SOT-MRAM based inference accelerators for binary neural networks. The architecture in \cite{MRAManalogPIM} utilizes analog peripheral circuits to achieve multi-bit computation. However, most of the previous SOT-MRAM based PIM accelerators either support only single bit computations or are too complex to be used for DNN training. A recent work \cite{SOT-MRAMbasic} introduces an efficient way to realize a complete set of Boolean logic functions, i.e., AND, OR, and XOR (see Figure~\ref{fig_logicfunc}) based on a single MTJ device, paving the way for SOT-MRAM based digital PIM accelerators.

\begin{figure}[!t]
\centering
\includegraphics[width=3.3in]{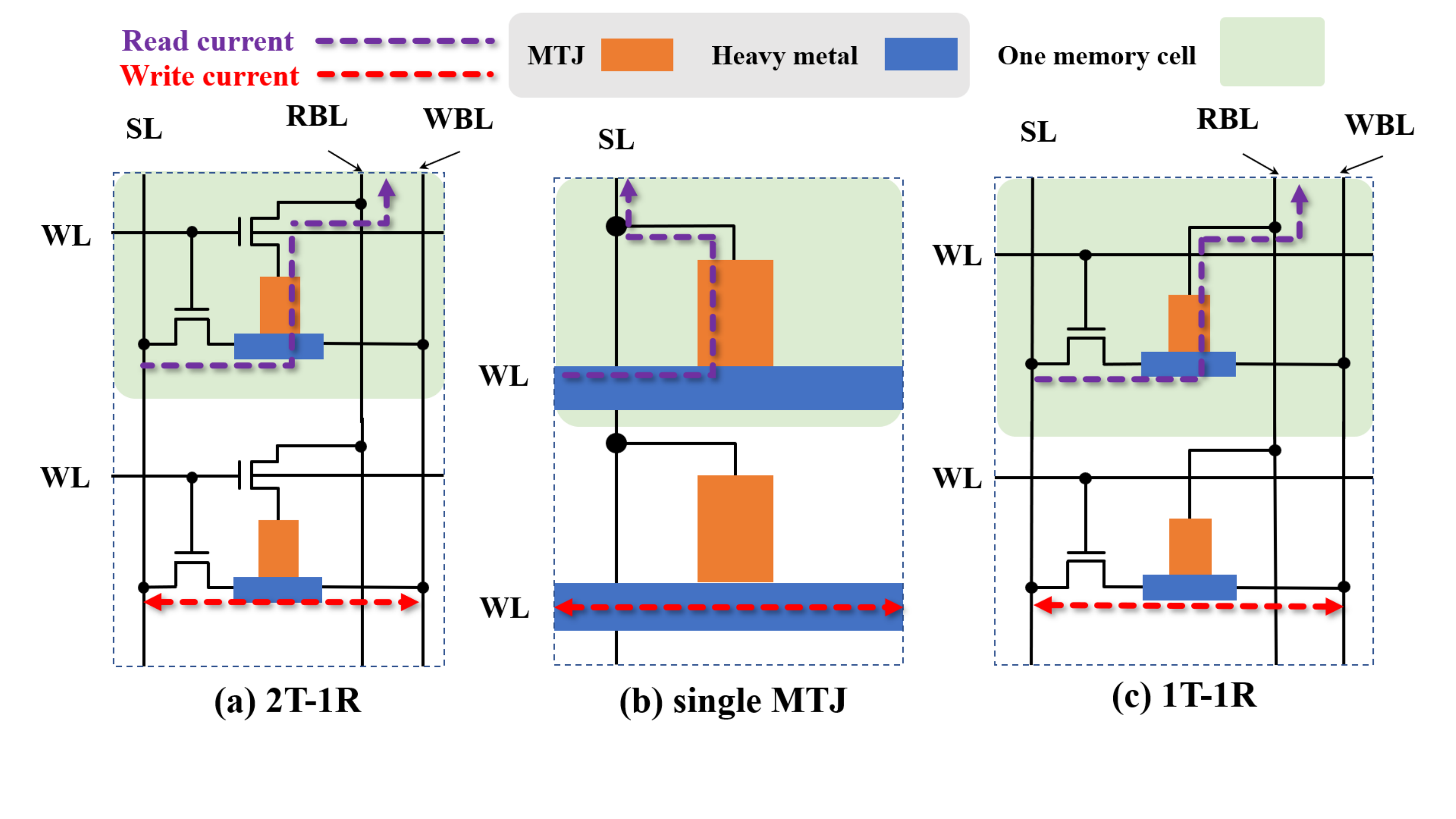}
\vspace{-2.0em}
\caption{The (a) 2T-1R \cite{SOT-MRAMbasic}, (b) single MTJ \cite{SOT-MRAMbasic}, and (c) proposed 1T-1R cells, where WL, SL, RBL, and WBL denote the word-line, source-line, read bit-line, and write bit-line, separately, and the purple/red dash-lines in the top/bottom cells show the read/write current direction.}
\label{fig_cellcompare}
\vspace{-1.2em}
\end{figure}

\textbf{SOT-MRAM Memory Cells.}
Figure~\ref{fig_cellcompare} (a) and (b) show two typical SOT-MRAM cell designs \cite{SOT-MRAMbasic}. 
The 2T-1R cell in Figure~\ref{fig_cellcompare} (a) consists of two transistors and one MTJ device.
During a read operation, a small negative voltage is applied on RBL and a read current between the read bit-line (RBL) and source-line (SL) is generated, enabling the data stored in the MTJ to be read out. During a write operation, a high voltage and a positive $V_b$ are applied to the selected word-line (WL) and RBL, respectively, generating a current between the SL and the write BL (WBL) to enable writing the target data into the MTJ. The voltage applied on RBL determines whether the corresponding device's resistance state can be switched. As a result, data can be written into different devices in the same row simultaneously. Compared to this 2T-1R cell, our proposed cell (see  Figure~\ref{fig_logicfunc} (c)) maintains its row-parallel writing flexibility, while having less transistors and higher memory density and write speed. 

The memory cell in Figure~\ref{fig_cellcompare} (b) consists of only a single MTJ device.
During a write operation, the transistor at the target row and column are selected while all other transistors are set to be off, and either a zero or $V_b$ is applied for each column depending on the desired data value to be written into the MTJ. During a read operation, the transistors for all of the columns of the selected row are activated while other transistors are set to be off, enabling all data stored in the selected row to be read out in parallel.
While this single MTJ device cell has reduced parasitics and improved memory density, the current direction of all the cells in one row has to be changed simultaneously, requiring one extra step (as compared to the 2T-1R cell) for a write operation and thus limiting the computational latency of this single MTJ device cell as write operations dominate in the read/write process.

\textbf{Available Boolean Functions vs. Computational complexity.}
\label{subsec:trade-off}
Digital PIM accelerators' computational complexity is closely related to the Boolean functions that can be implemented in the adopted memory technology. Specifically, if the memory technology can realize a complete set of Boolean functions, computational procedures will be simpler \cite{FloatPIM, drisa, SOT-MRAMbasic}; whereas more complicated procedures are required if only a limited set of Boolean functions can be implemented. As a concrete example, as the resistive random-access memory (ReRAM) in \cite{FloatPIM} supports only logic function NOR operations, it requires 13 steps of cell switch using a total of 12 cells for implementing 1-bit full addition (FA), while the same operation in SOT-MRAM based design requires only 5 steps of cell switch using a total of 4 cells \cite{SOT-MRAMbasic}. Note that the 1-bit FA in \cite{SOT-MRAMbasic} is not suitable for DNN training, because their design overwrites the
original operands which are still needed later during training.

\textbf{Floating Point Computation.} It is well recognized that training with floating
point precision favors high classification accuracy.
\textit{Addition:} among the standard procedures of performing floating point additions, it is naturally feasible to process all steps of different additions in parallel except for the step of aligning the two operands' exponents. This is because different additions might require different shifted bits for their exponent alignment. To efficiently handling this step, FloatPIM \cite{FloatPIM} processes all the mantissas that require the same shifted amounts in parallel. \textit{Multiplication:} similarly, the time/energy dominant step when performing floating point multiplication is the multiplication of the two operands' mantissas. For efficiently handling this step, FloatPIM \cite{FloatPIM} processes input-weight multiplications in a row-wise parallel manner, which however involves writing a large amount of memory cells (e.g., 455 cells at one row for a 32-bit multiplication) in order to store the intermediate results. As writing into a memory cell can cost 100$\times$ higher energy than that of a NOR operation \cite{FloatPIM}, new methods with much improved time/energy efficiency are needed.

\vspace{-0.6em}

\section{The Proposed Digital PIM Accelerator}
\vspace{-1em}


\subsection{A New 1T-1R Memory Cell}
\vspace{-1em}

Inspired by the advantages and disadvantages of the 2T-1R and single MTJ memory cells in \cite{SOT-MRAMbasic} and in order to harvest the benefits of both, we propose a new 1T-1R memory cell as shown in Figure~\ref{fig_cellcompare} (c) that consists of four control terminals. During both read and write operations, a high voltage (e.g., 0.7V in a 28nm technology) is applied to WL for turning on the selected transistor. \textit{During a read operation}, RBL and SL are applied with a small negative voltage (e.g., -100mV) and connected to the ground, separately, resulting in a current flowing from SL to RBL (see the purple dash-line in Figure~\ref{fig_cellcompare} (c)). Note that the negative voltage on RBL increases the current threshold to switch MTJs' resistance state in order to avoid undesirable switches when reading data \cite{MRAMparameters}. \textit{During a write operation}, a positive voltage $V_{b}$ (e.g., 600mV in \cite{MRAMparameters}) or 0 is applied to RBL for controlling the threshold of current switching, while a positive or negative voltage is formed between WBL and SL to (1) generate the write current and (2) control the current direction.  In this way, we can perform logic functions as shown in Figure~\ref{fig_logicfunc} in the write process. For example, the computation of the OR operation is as follows: considering $A=1$ (i.e., $V_b$ is applied to the top of SOT-MRAM), the write current flowing from SL to WBL (i.e., $C=1$) is larger than the threshold of current switching, leading to the MTJ's switching to a high resistance state, i.e., $B_{i+1}$=1.

Compared with the existing SOT-MRAM based memory cells, the proposed memory cells feature increased memory density and improved read speed (e.g., over the 2T-1R cell in Figure~\ref{fig_logicfunc}), while maintaining the capability to control different cells within the same row which enables high computational flexibility and reduced computational latency.



\vspace{-0.8em}
\subsection{A New FA for Digital PIM Accelerators}
\vspace{-1.1em}

As discussed in Section \ref{subsec:trade-off}, existing implementations of FA in digital PIM accelerators either require complicated procedures (and thus high energy/time cost) \cite{FloatPIM} or are not suitable for DNN training due to the overwriting of operands \cite{SOT-MRAMbasic}. To this end, we propose a new FA design that addresses the limitation of prior works. Mathematically, 1-bit FA can be expressed as,


\vspace{-1.3em}
\begin{equation}
\begin{array}{l}{S(R)=X \oplus Y \oplus Z} \\ {Z'=X Y+Z(X \oplus Y)}\end{array}
\end{equation}
\vspace{-1.3em}

\begin{figure}[!t]
\centering
\includegraphics[width=3.3in]{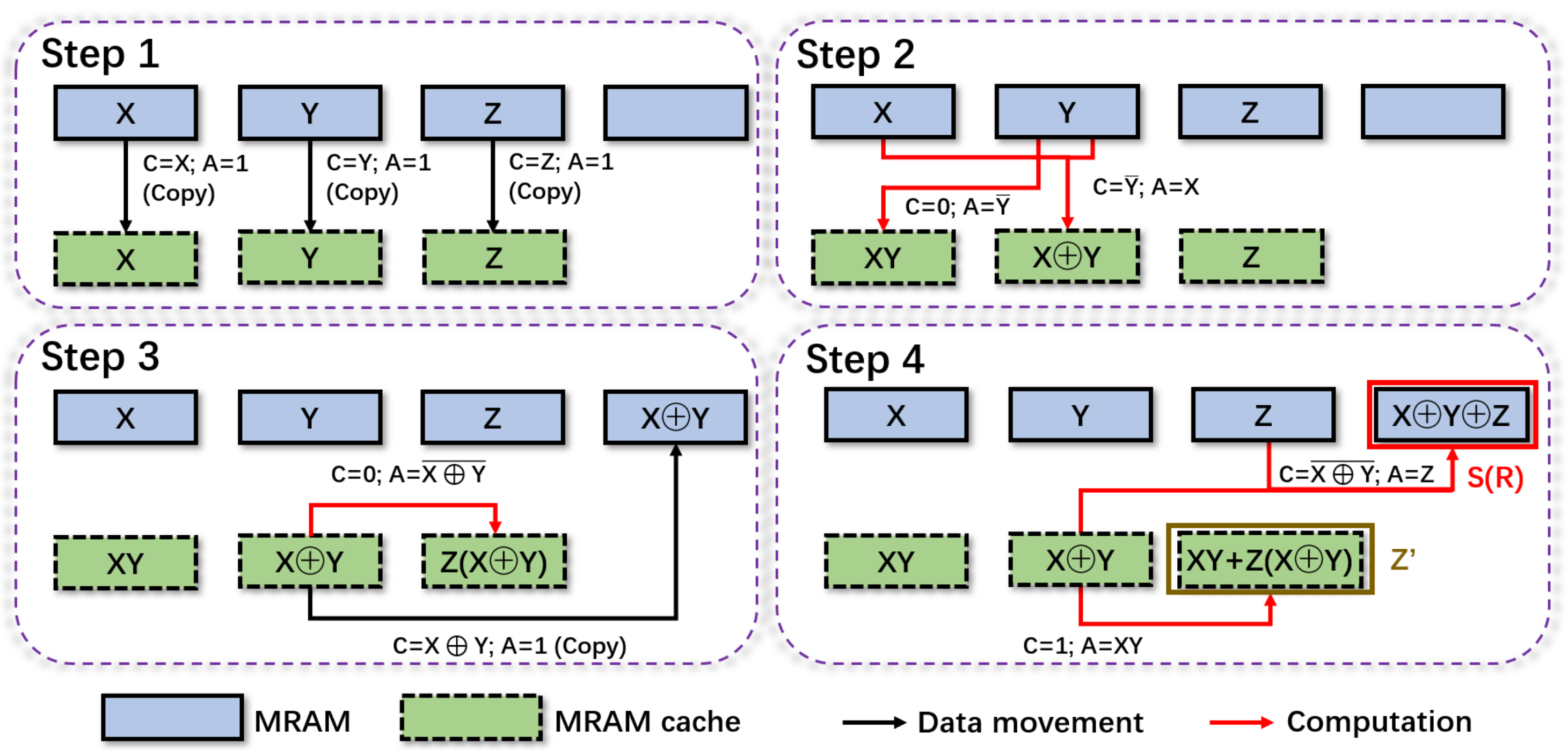}
\vspace{-0.5em}
\caption{The procedure of the proposed FA with each step features parallel read and then write.}
\label{fig_fulladd}
\vspace{-1.8em}
\end{figure}


where X and Y are two operands, Z is the input carry, and Z' is the output carry. Figure~\ref{fig_fulladd} shows the required procedure: 1) \textit{Step 1} - X, Y, and Z are copied to corresponding MRAM caches of the same columns; 2) \textit{Step 2} - Both an XOR and AND operations between X and Y are performed in parallel to obtain X$\oplus$Y and XY; 3) \textit{Step 3} - X$\oplus$Y is copied to the same row of and next to Z, and an AND operation between Z and X$\oplus$Y is performed; and 4) \textit{Step 4} - an XOR operation between Z and X$\oplus$Y is performed in parallel with an OR operation between XY and Z(X$\oplus$Y). These four steps of read and write operations result in the sum S(R) and the new carry Z', while the value and location of X and Y are kept unchanged. 

The aforementioned process can be performed using column-wise parallelism. Unlike \cite{FloatPIM}, the operands and results in our design can be assigned to different columns.  
The MRAM cache can be reused in sequential 1-bit full additions for multi-bit
additions. Specifically, our proposed FA requires 4 steps of read and write using a total of 4 memory cells, as compared to the required 13 equivalent steps using 12 memory cells in FloatPIM \cite{FloatPIM}. This improvement benefits from the : (1) available of complete Bolean functions in SOT-MRAM developed by \cite{SOT-MRAMbasic}, and (2) reuse of memory cells as cache and highly parallelable procedure.

\vspace{-0.8em}
\subsection{Floating Point Computation}
\vspace{-1em}
\textbf{Addition.} For handling the time/energy dominant exponent alignment, we adopt a similar ``search" method as \cite{FloatPIM} (see Section \ref{subsec:trade-off}), as shown in Figure~\ref{fig_floatcomp}(a). Specifically, if the $input$ bit matches with the stored bit, the read current from SL flows through the cell of a high resistance state and will be relatively low. Otherwise, the read current from SL flows through the cell of a low resistance state and will be relatively high. As a result, we can identify if the the $input$ bit matches with stored $exp^{'}$s (difference between two operands' exponents) based on the current flowing from SL. Consider $N_m$ bits for the mantissa and $N_e$ bits for the exponents. Unlike FloatPIM which only supports bit-by-bit shifting and requires exponent-alignment latency and energy proportional to $O\left({N_m}^2\right)$ \cite{FloatPIM}, the capability of shifting flexible bits thanks to the proposed 1T-1R cells enables an exponent-alignment latency and energy on the order of $O\left(N_m\right)$. Specifically, the latency and energy of our proposed floating point addition are:

\begin{figure}[!t]
\centering
\includegraphics[width=3.3in]{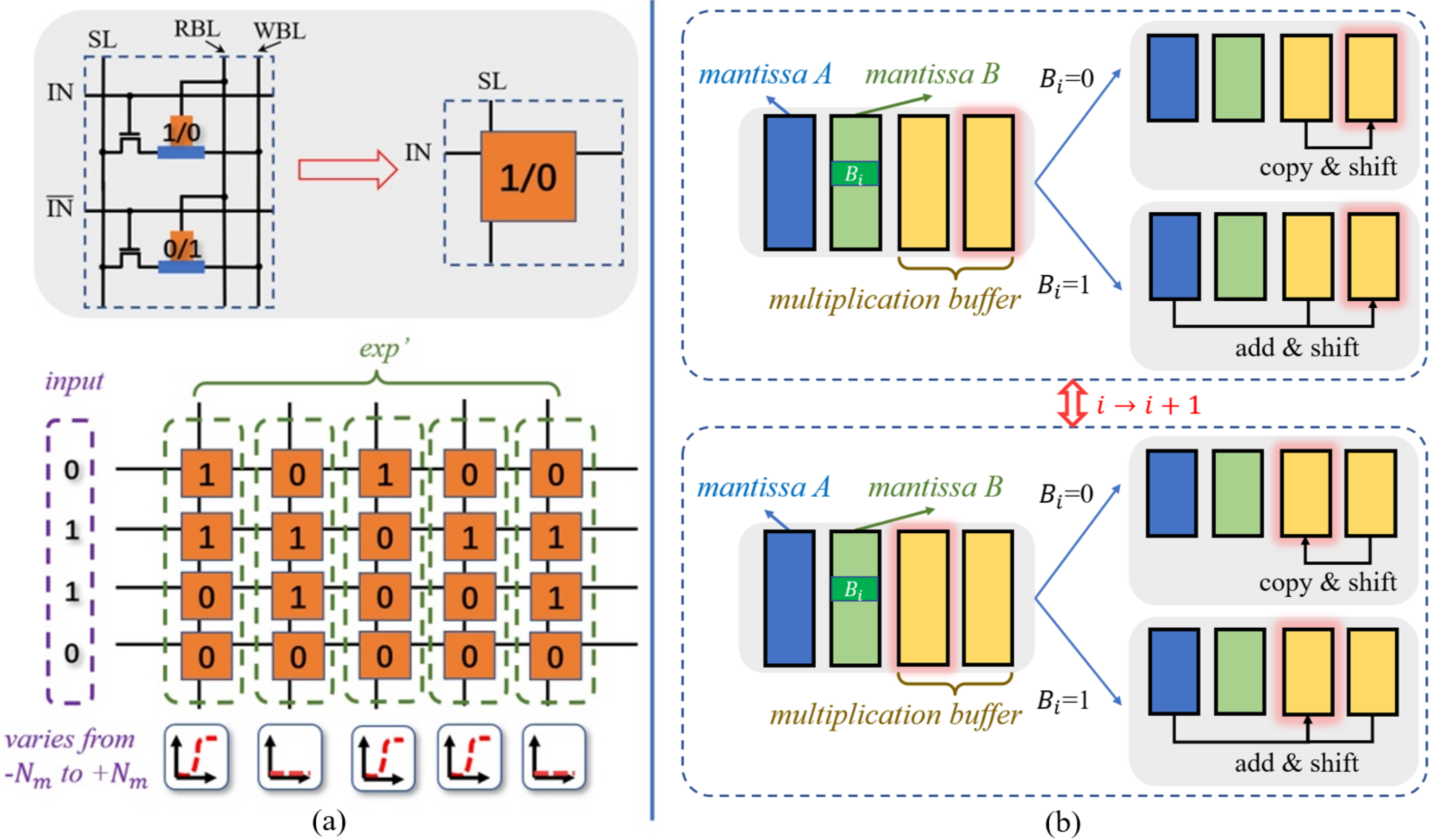}
\vspace{-0.5em}
\caption{The (a) ``search" method and (b) mantissa multiplication.}
\label{fig_floatcomp}
\vspace{-1.8em}
\end{figure}

\vspace{-0.8em}
\begin{equation}
\begin{aligned} T_{\text{add}}=(1+7 N_{e}+7 N_{m}) T_{\text{read}}+(7 N_{e}+7 N_{m}) T_{\text {write}} \\ +2(N_{m}+2) T_{\text {search}}  \\ E_{\text {add}}=(1+14 N_{e}+12 N_{m}) E_{\text {read}}+(14 N_{e}+12 N_{m}) E_{\text {write}} \\+2(N_{m}+2) E_{\text {search}} \end{aligned}
\label{eq:addition}
\end{equation}
\vspace{-0.8em}

\textbf{Multiplication.}
In a high-level, our proposed floating point multiplication (Figure~\ref{fig_floatcomp} (b)) involves only shift-and-add operations for the dominant mantissa multiplications. Specifically, the multiplicand is multiplied with a single bit of multiplier, shifted, and added to previous intermediate results add multiplier's bit times. The intermediate result of previous and current add are stored in two columns of cells, which will switch their roles in the next add operation. 
Other steps in multiplication are performed using column-wise AND, OR, and XOR operations in 1T-1R cells. The  multiplication latency and energy costs are:

\vspace{-0.8em}
\begin{equation}
\begin{array}{l}{T_{\text{mul}}=(2 N_{m}^{2}+6.5 N_{m}+6 N_{e}+3)(T_{\text{read}}+T_{\text{write}})} \\ {E_{\text{mul}}=(4.5 N_{m}^{2}+11.5 N_{m}+13.5 N_{e}+6.5)(E_{\text {read}}+E_{\text {write}})}\end{array}
\label{eq:mul}
\end{equation}
\vspace{-0.5em}


 \vspace{-2em}
\section{Experiment Results}
\vspace{-1em}
\subsection{Experimental Methodology and Setup}
\vspace{-1em}


\textbf{Methodology:} To evaluate the effectiveness of the proposed ideas, 
we integrate the aforementioned 1T-1R SOT-MRAM memory cell, new FA, and efficient floating point computation to realize an SOT-MRAM based digital PIM accelerator for DNN training. For bench-marking the resulting accelerator in terms of DNN training performance, we first obtain (1) the energy consumption and latency per one bit memory write/read operation and per multiplication/addition calculation (MAC) of 32-bit floating point precision (commonly used for DNN training) and (2) the area of memory array and peripherals, separately; and then compare the training performance of our proposed accelerator with a state-of-the-art design, FloatPIM in \cite{FloatPIM} based on LeNet-type DNN model with 21,690 parameters of 32-bit floating point precision. Note that computations in both designs are performed with full precision, resulting in the same test accuracy after training. 

\textbf{Setup:} To estimate the energy and latency cost per one bit memory write/read and the area in the proposed accelerator, we incorporate (1) basic SOT-MRAM cell parameters (see Table~\ref{table_parameter}) from~\cite{MRAMparameters} and (2) the current sense amplifier in~\cite{senseamplifier} into the state-of-the-art simulator NVSim in~\cite{NVSim}; while the energy and latency consumption per one bit memory write/read and the area 
of the FloatPIM accelerator is obtained from their paper \cite{FloatPIM}. For evaluating the accelerator performance, we adopt the same memory subarray size of 1024$\times$1024 and hardware architecture as the FloatPIM baseline \cite{FloatPIM} for a fair comparison. The performance of both the proposed and FloatPIM accelerators are then obtained by designing a dedicated PIM accelerator simulator, the estimated performance of which is validated to be consisted with ($<$10\% prediction accuracy) the reported performance in \cite{FloatPIM} under various conditions.



\begin{table}[!t]
\vspace{-0.3em}
\renewcommand{\arraystretch}{1.5}
\caption{Parameters of a SOT-MRAM cell\cite{MRAMparameters}}
\label{table_parameter}
\vspace{-0.3em}
\centering
\begin{tabular}{|c||c||c||c||c||c|}
\hline
$R_{\text{on}}$ & $R_{\text{off}}$ & $V_{b}$ & $I_{\text{write}}$ & $t_{\text{switch}}$ & $E_{\text{switch}}$\\
\hline
50 $k\Omega$ & 100 $k\Omega$ & 600 mV & 65 $\mu$A & 2.0 ns & 12.0 fJ\\
\hline
\end{tabular}
\vspace{-1.2em}
\end{table}

\vspace{-1em}
\subsection{MAC with Floating Point Precision}\label{sec:MAC}
\vspace{-1em}

\begin{figure}[!t]
\vspace{-1em}
\centering
\includegraphics[width=3.4in]{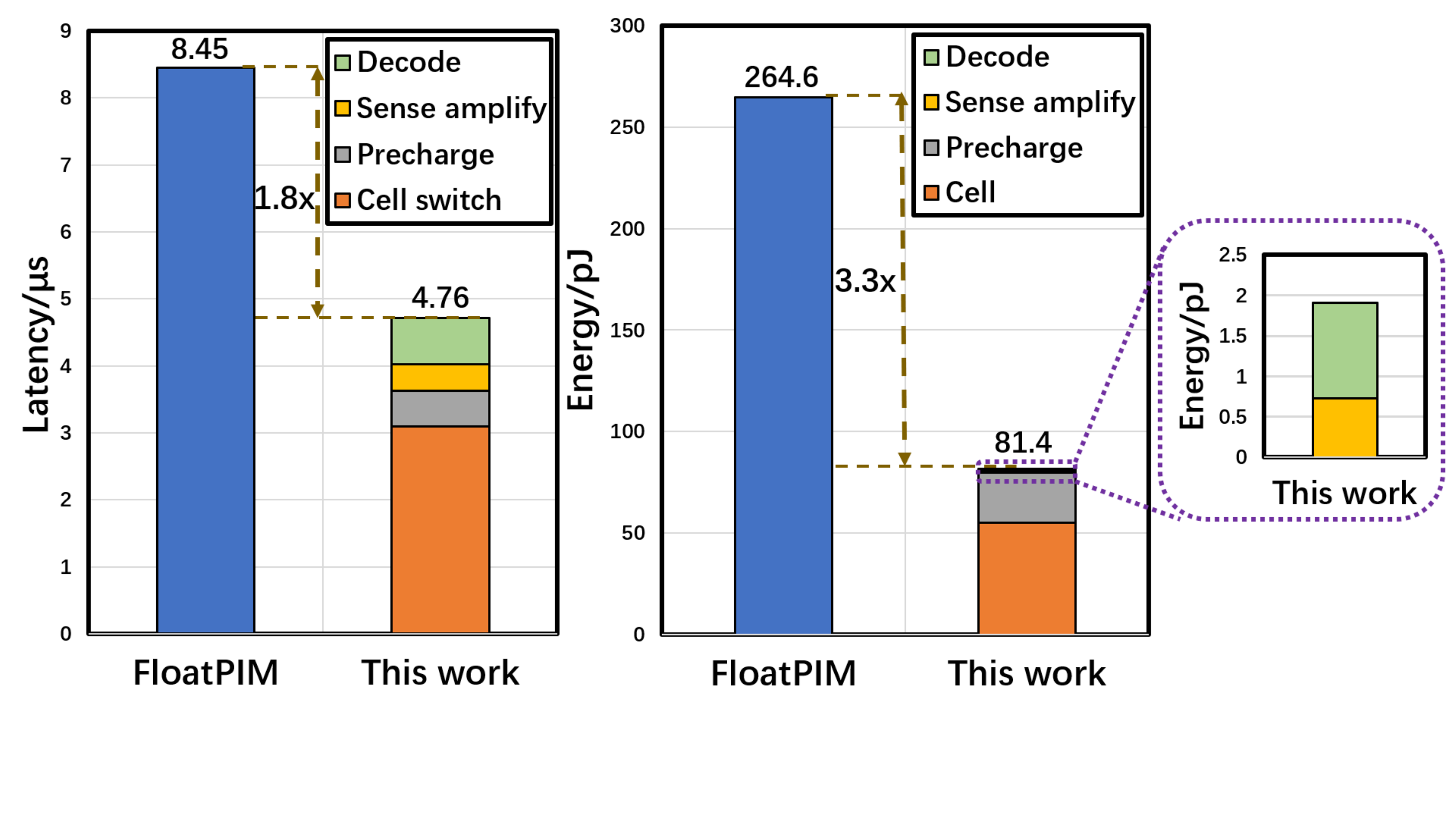}
\vspace{-3.5em}
\caption{Comparing the proposed MAC with that of FloatPIM in terms of latency (left) and energy (right) cost, where our design's latency and energy breakdown is also shown.}
\label{fig_MACcompare}
\end{figure}

Figure~\ref{fig_MACcompare} evaluates the energy cost and latency of a MAC using the proposed 1T-1R cell, FA, and floating point addition and multiplication based on a 1024$\times$1024 subarray. We can see that our MAC achieves a 3.3$\times$ lower energy cost and 1.8$\times$ lower latency compared with FloatPIM when finishing the same computation in the same size subarray, thanks to the facts that (1) the required fewer computing steps (i.e., read and write operations) to perform the same computation; and (2) compared with ReRAM in FloatPIM, the adopted SOT-MRAM requires a lower write current and thus a lower energy cost and latency for precharge. In addition, Figure \ref{fig_MACcompare} shows that cell switch latency dominates a MAC's latency. Fortunately, ultra-fast switching SOT-MRAM developed recently\cite{fastMRAM} can potentially further reduce our MAC's latency. Specifically, if we use the switch time in \cite{fastMRAM} to replace the current one, the MAC latency will be reduced by 56.7$\%$.

\vspace{-1em}
\subsection{Training with Floating Point Precision}\label{sec:training}
\vspace{-1.2em}

\begin{figure}[!t]
\vspace{-1em}
\centering
\includegraphics[width=3.3in]{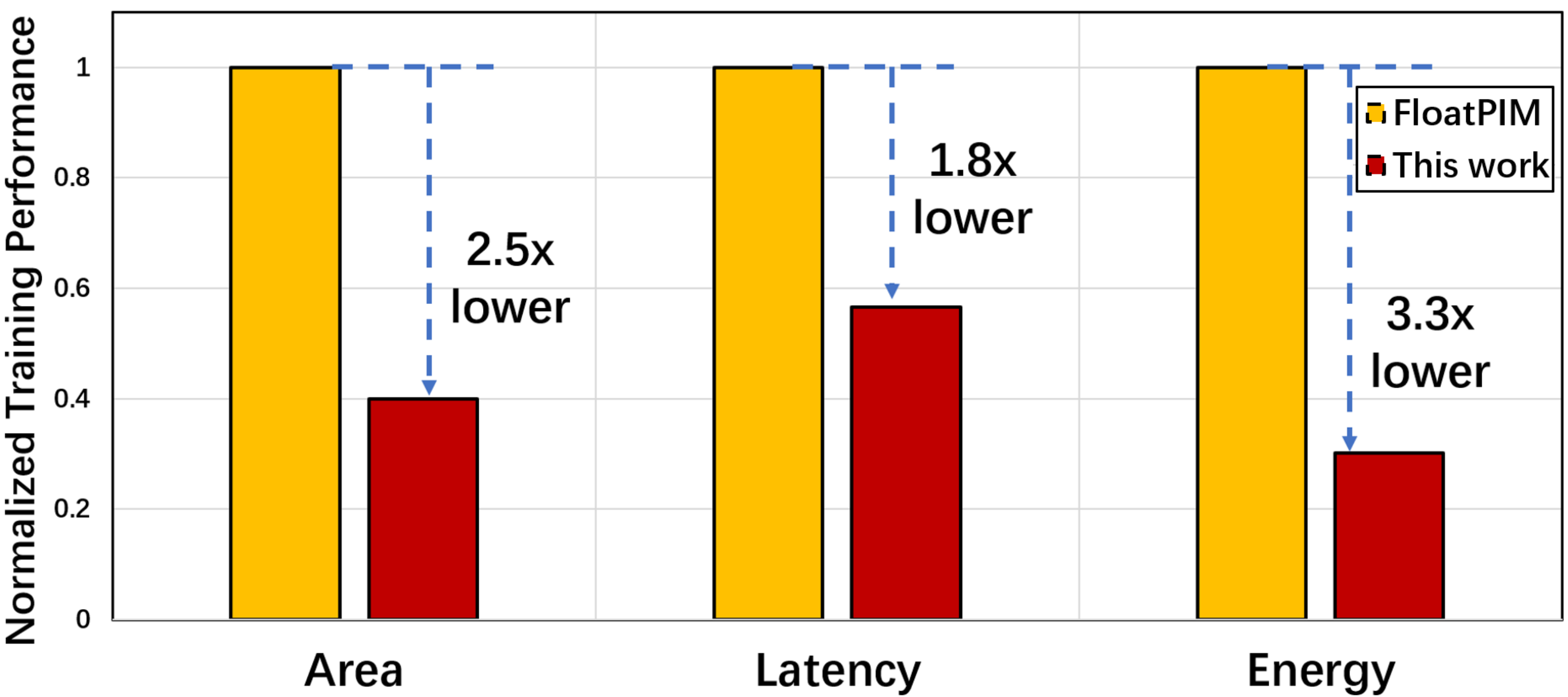}
\vspace{-0.5em}
\caption{The training performance of the proposed accelerator normalized over FloatPIM in terms of area, latency, and energy cost based on a LeNet-5 model and the MNIST dataset with a test accuracy of 97.08$\%$.}
\label{fig_training}
\vspace{-1em}
\end{figure}



Based on the simulation results of NVSim, we obtain the total energy cost, latency, and required area of training the LeNet model based on MNIST 
using the proposed digital PIM accelerator as benchmarked over FloatPIM in Figure~\ref{fig_training}. It is shown that our accelerator achieves a 2.5$\times$, 1.8$\times$, and 3.3$\times$ lower area, latency, and energy consumption, respectively, as compared to FloatPIM. This improved performance is resulted from the facts that (1) the fewer cells required to store intermediate results when performing the same computation as described in Section 3 and Figure~\ref{fig_fulladd}; and (2) the higher design flexibility: unlike FloatPIM which requires the operands, intermediate results and the finial result must be stored in the same row when performing addition or multiplication, memory cells in the proposed accelerator can be used to store intermediate results flexibly which maximizes cell reuse opportunities. Note that the improvement of the proposed accelerator's energy efficiency and latency over FloatPIM is similar to that of a MAC, because computation dominates the total energy consumption and latency of small LeNet training.

\vspace{-1em}
\section{Conclusion and future work}
\vspace{-1em}
In this paper, we propose a SOT-MRAM based digital PIM accelerator for DNN training. Comparing to the pioneering work of ReRAM based digital PIM accelerator FloatPIM\cite{FloatPIM}, our design achieves 3.3$\times$ higher energy efficiency, 1.8$\times$ lower latency, and 2.5$\times$ higher area efficiency. Future work will investigate the scalability of the proposed accelerator and evaluate the proposed design in larger DNN models and dataset. 

\section{Acknowledgement}
\vspace{-1em}
The work is supported by the National Science Foundation (NSF) through the ECCS Division Of Electrical, Communication \& Cyber System (Award number: 1934767).

\bibliographystyle{IEEEtran}

\end{document}